\documentclass[aps,prl,twocolumn,showpacs,superscriptaddress,groupedaddress]{revtex4-1}  
\usepackage[dvipdfmx]{graphicx}  
\usepackage{dcolumn}   
\usepackage{bm}        
\usepackage{amssymb}   
\usepackage{natbib}
\begin{document}


\setlength{\oddsidemargin}{-0.7cm}
\setlength{\evensidemargin}{0.0cm}
\setlength{\textwidth}{17.5cm}
\makeatletter
  \newcounter{subeqncnt}
  \def\thesubeqncnt{\alph{subeqncnt}}
  \def\subequations{\begingroup%
     \stepcounter{equation}\edef\@tempa{\theequation}%
     \let\c@equation\c@subeqncnt\c@subeqncnt\z@
     \edef\theequation{\@tempa\noexpand\thesubeqncnt}}
  \let\endsubequations\endgroup
  \makeatother

\title{Size distribution of islands according to 2D growth model\\ 
with 2 kinds of diffusion atoms
}

\author{R. Yamauchi$^1$, X.M. Lu$^{1*}$, M. Koyama$^1$, H. Sasakura$^{1,2}$, Y. Nakata$^1$, and S. Muto$^1$\\
$^1$Department of Applied Physics, Graduate School of Engineering, Hokkaido University, Sapporo, Hokkaido 060-8628, Japan\\
$^2$Creative Research Institution, Hokkaido University, Sapporo, Hokkaido 001-0021, Japan\\
$^*$Present address: Center for Frontier Research of Engineering, Institute of Technology and
Science, Tokushima University, Tokushima 770-8506, Japan
}

\date{\today}

\begin{abstract}
We simulated the growth of 2D islands with 2 kinds of diffusion atoms using the kinetic Monte-Carlo (kMC) method. As a result, we found that the slow atoms tend to create nuclei and determine the island volume distribution, along with additional properties such as island density. We also conducted a theoretical analysis using the rate equation of the point-island model to confirm these results.
\end{abstract}

\maketitle


Nucleation is an important step in several phenomena, such as growth of clouds, polymers, and crystals. In many cases, heterogeneous nucleation occurs on surfaces and at interfaces, and it is also facilitated by the presence of impurities. The heterogeneous two-dimensional (2D) nucleation is particularly important because of its relevance to the epitaxial growth of single crystal semiconductors, of which quantum dots (QDs)\cite{b1,b2,b3,b4,b5} are one of the most attractive classes.

Many theoretical and experimental works have been conducted on the 2D nucleation and island size-distribution\cite{b6,b7,b8,b9,b10} However, most of the studies only dealt with one kind of diffusion atoms. We have investigated the volume distribution of QDs grown by molecular beam epitaxy (MBE), and we are now interested in ternary alloy semiconductor QDs, especially in InAlAs QDs, which have a luminescence wavelength of 750 nm, at which the Si photodetectors exhibit the maximum sensitivity\cite{b2,b3}. InAlAs QDs are ternary QDs, to which two group I\hspace{-.1em}I\hspace{-.1em}I adatoms, In and Al, with extremely different surface diffusion rates, contribute. We previously found that investigating the submonolayer 2D island growth provides helpful knowledge on the QD growth regime\cite{b4,b5}, and the 2D island growth by two kinds of largely different adatoms is of academic interest. In this paper, we examine the 2D nucleation and island size-distribution using both computer simulations and theoretical analysis of the point-island model. We found, for the first time, that the adatom with the slower diffusion tends to form nucleation centers and determines the size distribution of the islands.

First, we modeled 2D island growth using the kinetic Monte Carlo Method \cite{b7,b8}. In this model, we simulate the motion of atoms on a $400\ \times\ 400$ square lattice with a periodic boundary condition. At each stage, a new adatom is deposited on a randomly chosen site with a given deposition rate, $F$, or a previously-deposited adatom is chosen to diffuse to its nearest-neighbor site with a specific diffusion rate, $D$. At nucleation, the important parameter is the critical island size, $i$, such that $i+1$ aggregate atoms form a nucleus, which is a minimum island. Here we used $i=1$, which means that, if an adatom encounters another adatom as a neighbor, they are frozen at their sites and form a new island, resulting in nucleation. These processes are repeated up to a fixed coverage, $\theta$, which is the ratio of the number of all deposited atoms against the lattice area; typically, we set $\theta=0.3$. Here, $R = D/F$ characterizes the simulation for a fixed $\theta$. We then averaged the data from approximately 100 simulation runs. 
 
Regarding InAlAs QD growth, Al has a much smaller diffusion coefficient than In, because of the stronger Al-As bond in comparison with the In-As bond \cite{b11}. Therefore, we included two adatoms with different diffusion rates, $D_{f}$ and $D_{s}$, corresponding to fast In and slow Al, respectively. We varied $R_f = D_{f} /F$ and $R_{s} = D_{s} /F$ and observed their effects on the island size distribution. The flux ratio of these two adatoms represents the alloy composition of the dots, for which we used a value corresponding to the ${\rm In:Al}=0.7:0.3$ ratio used in our previous experiment \cite{b2}.

In a submonolayer epitaxial growth process such as that of homoepitaxial Fe/Fe(100) \cite{b9} or heteroepitaxial Fe/Cu(100) \cite{b10}, and also the InAs/GaAs QD used in our study \cite{b4}, the volume distribution of islands is known to show the scaling property
\begin{equation}
N_s=\theta S^{-2}f_i\left(\frac{s}{S}\right),
\end{equation} where $N_s$ is the density of an island of size $s$, $S$ is the average size, and $f_i(s/S)$ is the scaling function, which varies with the critical island size, $i$.

Fig. 1(a) shows the scaling plot at $ R_{f} =10^9$ with varying $R_s$. Each trendline has only a single peak, which translates leftward with decreasing $ R_{s} $. On the right side of the plot, we can see a tail larger than the emprial curve for a single atom\cite{b8}, illustrated by the solid curve. This means that these island distributions experience size dispersion in the case of large islands, where $ s/S>2 $. Here, we note that this feature is similar to our previous experiments on InAlAs QDs \cite{b3}. 

 \begin{figure}[b]
   \centering
    \includegraphics[clip,width=80mm]{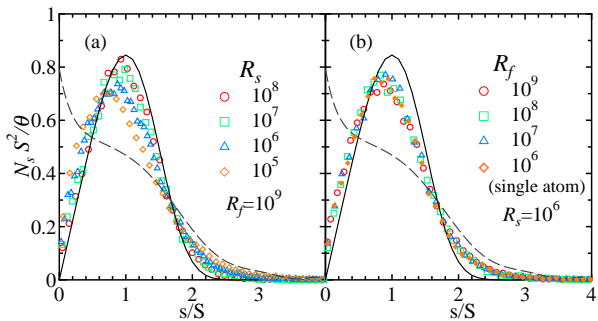}
   \caption{The island size scaling function with (a) varied $R_s$ and fixed $R_f$, and (b) varied $R_f$ and fixed $R_s$. They are compared with scaling functions for a single atom $i=1$ (solid line) and $i=0$ (dashed line). }
   \label{Fig:one}
  \end{figure}
 
Fig. 1(b) shows the scaling plot at $ R_{s} =10^6$ with varying $ R_{f} $. All curves are the same and this insensitivity to changes in $ R_{f} $ is also found for other $ R_s $ values. The scaling function is also equal to the single atom case with single $ R= R_{s} $,illustrated by the $R_f= 10^6$ result. These results imply that $R_{s}$ determines the scaling function regardless of the value of $R_{f}$.

For the scaling arguments the following is assumed
\begin{eqnarray}
N\sim R^{-\chi}\theta^{1-z}\ ,\\
N_1\sim R^{-\gamma}\theta^{-\nu}\ ,
\end{eqnarray} where $N=\sum_{s\geqq2}N_s$ is the total island density, $N_1$ is the monomer density, and $\chi,z,\gamma,$ and $\nu$ are critical exponents. 

\begin{figure}[t]
  \centering
   \includegraphics[clip,width=80mm]{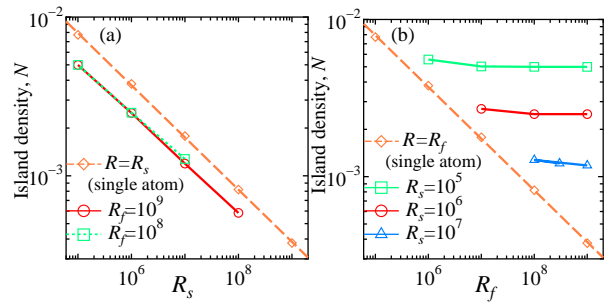}
  \caption{The island density for (a) fixed $R_s$ and varying $R_f$, and (b) fixed $R_f$ and varying $R_s$. The dashed lines show the case of single atom.}
  \label{Fig:two}
 \end{figure}
 
Fig. 2 shows the island density for cases with (a) varying $R_s$ and (b) varying $R_f$. As shown in Fig. 2(a), the dependence of the density on $ R_{s} $ is similar to the behavior of a single atom, and is expressed as $ N\approx R^{-\chi_i} $ or $ N\approx R^{-\chi_i}_s $ for fixed $\theta$. According to \cite{b12}, $\chi_i={i}/({i+2})$ and $\chi_1=1/3$ (for $i=1$), and our results give $\chi_1\approx0.33$ for the single atom case, and $\chi_1\approx0.31$($R_f=10^9$), $0.30$($R_f=10^8$) for cases featuring two atoms. On the other hand, as shown in Fig. 2(b), the density is almost independent of $ R_{f} $ for $R_{f}\gg R_{s}$. Therefore we can say that the slow adatom plays almost the same role about nucleation as the only diffusion atom in the single atom case.  

\begin{figure}[b]
 \centering
  \includegraphics[clip,width=80mm]{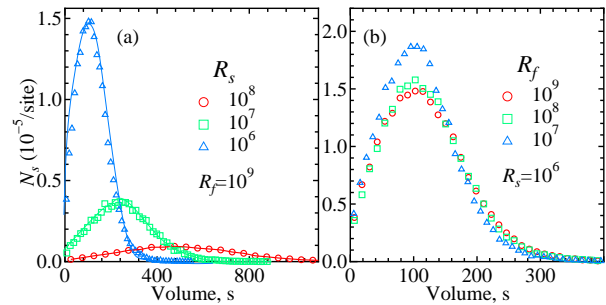}
 \caption{The "unscaled" island size distribution with (a) varied $R_s$ and fixed $R_f$, and (b) varied $R_f$ and fixed $R_s$.}
 \label{Fig:three}
\end{figure}

Fig. 3(a)  and Fig. 3(b) show the "unscaled" island size distribution with fixed $ R_{f} $ and fixed $ R_{s} $, respectively. As shown in Fig. 3(a), the distribution strongly depends on $ R_{s} $. This behavior is similar to that of a single atom with R. On the other hand, as shown in Fig. 3(b), the distributions are almost the same. This result inplies previous two features with  fixed $R_s$, same scaling functions in Fig.1(b) and same island densities in Fig.2(b). The slow adatom determines not only the scaling function but also the unscaled island size distribution itself.

Fig. 4(a) shows the origins of nucleation, $\emph{i.e.}$, the nucleus densities of islands composed of pairs of fast-fast, fast-slow, and slow-slow atoms. At the early stage, or for the cases of small coverage shown in the inset of Fig. 4(a), the number of nuclei containing slow atoms does not exceed the number of fast-fast nuclei. However, in the late stage, there are much more slow-slow nuclei than fast-fast nuclei, and even more fast-slow nuclei. At the same diffusion rate limit, $\emph{i.e.}$, $ R_{f} = R_{s} $, the number of nuclei containing slow adatoms, both slow-slow and fast-slow combinations, must be $1-(0.7)^2$, or 51\% of all nuclei in our $0.3:0.7$ composition. Here however, this ratio is over 90\% during the late stage. This demonstrates that slow monomers tend to be used as nuclei. This feature is clearly seen in Fig.4(b).

Fig. 4(b) shows monomer density for each kind of adatom. Except for very early stage, the slow monomer density is much larger than that of the fast monomers. Even in later stage in this simulation, when there are few or no monomers, the density of slow monomers is about $10^{-6}$ atoms/site, and that of fast monomers is about $10^{-7}\sim10^{-8}$ atoms/site on time average. These show that nucleation by slow adatoms originates from the large number of slow adatoms remaining on the surface.
\begin{figure}[t]
 \centering
  \includegraphics[clip,width=80mm]{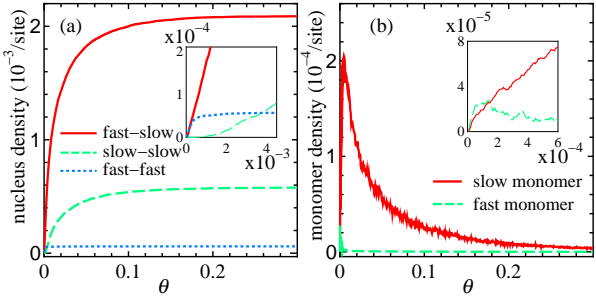}
 \caption{(a) Nuclei density for each atom pair, (b) monomer density for each atom composition, with $R_f=10^9$, $R_s=10^6$. The insets magnify the early stage.}
 \label{Fig:four}
 \end{figure}
 
To confirm the above finding, we also studied the kinetics of island growth processes theoretically, using the rate equation approach \cite{b7}. The rate equation describes the evolution of island size using the mean-field theory, which we extended to growth featuring 2 types of adatoms. Then, the fast monomer, $ N_{1}^f $, slow monomer, $ N_{1}^s $, and island with $s$ atoms, $N_s$, densities can be written as\begin{subequations}
\begin{eqnarray}
\frac{dN_1^f}{dt}&=& F_{f}-2K^f_1(N_1^f)^2-K^f_1N_1^fN_1^s\nonumber\\
&-& K^s_1N_1^sN_1^f-N_1^f\sum_{s=2}K^f_sN_s\ , \\
\frac{dN_1^s}{dt}&=& F_s-K^s_1N_1^sN_1^f-2K^s_1(N_1^s)^2\nonumber\\
&-& K^f_1N_1^fN_1^s-N_1^s\sum_{s=2}K^s_sN_s\ , \\
\frac{dN_s}{dt}&=& N_1^f(K^f_{s-1}N_{s-1}-K^f_sN_s)\nonumber\\
&+& N_1^s(K^s_{s-1}N_{s-1}-K^s_sN_s)\ , 
\end{eqnarray}
\end{subequations}where $F_\alpha$ is the deposition rate of adatom $\alpha$ ($\alpha=f, s$, and $F_{f}+F_{s}= F$), and $K_s^\alpha$ is the rate of adatom $\alpha$ attachment to islands of size $s$. We assume that $K^\alpha_s\sim D_\alpha s^p$, $ D_\alpha $ is the diffusion rate of adatom $ \alpha $, and the exponent, $p$, differs by the shape or dimension of the islands. Dividing the equations by $ F $, they can then be rewritten as
\begin{subequations}
\begin{eqnarray}
\frac{dN_1^f}{d\theta}&=&f_f-2R_f(N_1^f)^2-R_fN_1^fN_1^s\nonumber\\
&-&R_sN_sN_1^f-R_fN_1^f\sum_{s=2}s^pN_s\ , \\
\frac{dN_1^s}{d\theta}&=&f_s-R_sN_1^sN_1^f-2R_s(N_1^s)^2\nonumber\\
&-&R_fN_1^fN_1^s-R_sN_1^s\sum_{s=2}s^pN_s\ ,\\
\frac{dN_s}{d\theta}&=&[R_fN_1^f+R_sN_1^s]\nonumber\\
&\times&[(s-1)^pN_{s-1}-s^pN_s]\ ,
\end{eqnarray}
\end{subequations}where $R_\alpha=D_\alpha/F$, $ \theta=Ft$ is the coverage, and $f_\alpha=F_\alpha/F$ is the deposition ratio of adatom $ \alpha $. Rewriting in terms of the scaled variables: $\hat{\theta}=R^{1/2}_{f}\theta$, $\hat{N_s}=R_f^{1/2}N_s$, and $\hat{N_1^\alpha}=R^{1/2}_{f}N_1^\alpha$, the equations become
\begin{subequations}
\begin{eqnarray}
\frac{d\hat{N}_1^f}{d\hat{\theta}}&=&f_f-2(\hat{N}_1^f)^2-\hat{N}_1^f\hat{N}_1^s\nonumber\\
&-&r\hat{N}_1^s\hat{N}_1^f-\hat{N}_1^f\sum_{s=2}s^p\hat{N}_s\ , \\
\frac{d\hat{N}_1^s}{d\hat{\theta}}&=&f_s-r\hat{N}_1^s\hat{N}_1^f-2r(\hat{N}_1^s)^2\nonumber\\
&-&\hat{N}_1^f\hat{N}_1^s-r\hat{N}_1^s\sum_{s=2}s^p\hat{N}_s\ , \\
\frac{d\hat{N}_s}{d\hat{\theta}} 
&=&[\hat{N}_1^f+r\hat{N}_1^s][(s-1)^p\hat{N}_{s-1}-s^p\hat{N}_s]\ ,
\end{eqnarray}
\end{subequations} where $r=R_{s}/R_{f}\ll1$. 

 We further assumed the point-island model \cite{b6}, where each island occupies only one site. Then, $K_\alpha\sim D_\alpha$ is independent of $s$ $(p=0)$, and $\sum_{s=2}s^p\hat{N}_s=\sum_{s=2}\hat{N}_s=\hat{N}.$ Hence,
 \begin{subequations}
 \begin{eqnarray}
 \frac{d\hat{N}_1^f}{d\hat{\theta}}&=&f_f-2(\hat{N}_1^f)^2-\hat{N}_1^f\hat{N}_1^s\nonumber\\
 &-&r\hat{N}_1^s\hat{N}_1^f-\hat{N}_1^f\hat{N}\ , \\
 \frac{d\hat{N}_1^s}{d\hat{\theta}}&=&f_s-r\hat{N}_1^s\hat{N}_1^f-2r(\hat{N}_1^s)^2\nonumber\\
 &-&\hat{N}_1^f\hat{N}_1^s-r\hat{N}_1^s\hat{N}\ , \\
 \frac{d\hat{N}}{d\hat{\theta}} 
  &=&\sum_{s=2}\frac{d\hat{N_s}}{d\hat{\theta}}=[\hat{N}_1^f+r\hat{N}_1^s][\hat{N}_1^f+\hat{N}_1^s]\ .
 \end{eqnarray}
 \end{subequations}
 
 In the early stages (low coverage, $\hat{\theta}\ll1$), the number of monomers is quite low, and which is larger than that of islands, so that $\hat{N}\ll\hat{N}_1^{\alpha}$. Therefore, the last 4 terms in Eqs. 7(a) and 7(b) can be neglected, and therefore
 \begin{subequations}
 \begin{eqnarray}
 N_1^f&=&f_f\theta\, \\
 N_1^s&=&f_s\theta\, \\
 N&\sim&(f_f+rf_s)R_f\theta^3\sim f_fR_f\theta^3\ .
 \end{eqnarray}
 \end{subequations} In later stages ($\hat{\theta}\gg1$), the island density has increased while the monomer density has decreased due to shortened diffusion length. Hence, $\hat{N}\gg\hat{N}_1^\alpha$ and  ${d\hat{N}^{\alpha}_1}/{d\hat{\theta}}\ll f_{\alpha}$. Therefore,
 \begin{subequations}
 \begin{eqnarray}
  \frac{d\hat{N}_1^f}{d\hat{\theta}}&\sim&f_f-\hat{N}_1^f\hat{N}\simeq 0\ , \\
  \frac{d\hat{N}_1^s}{d\hat{\theta}}&\sim&f_s-r\hat{N}_1^s\hat{N}\simeq 0\ , \\
  \frac{d\hat{N}}{d\hat{\theta}} 
  &\sim&[\hat{N}_1^f+r\hat{N}_1^s][\hat{N}_1^f+\hat{N}_1^s]\ .
  \end{eqnarray}\\
  \end{subequations} Using the condition $r\ll1$, $f_{f}+f_{s}/r=f_{s}/r$, the solution can be rewritten as
\begin{subequations}
\begin{eqnarray}
N_1^f&\sim&rf_ff_s^{-\frac{1}{3}}R_s^{-\frac{2}{3}}\theta^{-\frac{1}{3}}\ , \\
N_1^s&\sim&f_s^{\frac{2}{3}}R_s^{-\frac{2}{3}}\theta^{-\frac{1}{3}}\ , \\
N&\sim&f_s^{\frac{1}{3}}R_s^{-\frac{1}{3}}\theta^{\frac{1}{3}}\ .
\end{eqnarray}
\end{subequations}Apart from the factor of $ f_\alpha $, $N$ is equal to the single atom case solution (replacing $ R $ with $ R_{s} $), $N\sim R^{-1/3}\theta^{1/3}$[6],  in accordance with Fig. 2. In addition, the property $ N^f_1\sim rN_1^s\ll N_1^s $ corresponds to Fig. 4(b). Moreover, the fact that the island distribution $N_s$ is independent of $R_f$, seen in Figs. 1(b) and 3(b), can easily be obtained by using eqs. (5c) and (10a) -(10c) with $p=0$ iteratively.

Also, we can discuss the contents of the nuclei $N^{f-f},N^{f-s},$ and $N^{s-s}$, as shown in Fig. 4 (a). Here $N^{\alpha-\beta}$ represents the density of islands which have nucleated using atoms $\alpha$ and $\beta$ ($\alpha ,\beta =f\ or\ s$, arbitrary order). The rate equations of these densities can be written as
\begin{subequations}
\begin{eqnarray}
\frac{dN^{f-f}}{dt}&=&K^f_1(N_1^f)^2\ , \\
\frac{dN^{f-s}}{dt}&=&K^f_1N_1^fN_1^s+K^s_1N_1^sN_1^f\ , \\
\frac{dN^{s-s}}{dt}&=&K_1^s(N_1^s)^2\ .
\end{eqnarray}
\end{subequations}These can be rewritten, using the point-island model, as
\begin{subequations}
\begin{eqnarray}
\frac{d\hat{N}^{f-f}}{d\hat{\theta}}&=&(\hat{N}_1^f)^2\, \\
\frac{d\hat{N}^{f-s}}{d\hat{\theta}}&=&(1+r)\hat{N}_1^f\hat{N}_1^s\sim \hat{N}_1^f\hat{N}_1^s\, \\
\frac{d\hat{N}^{s-s}}{d\hat{\theta}}&=&r(\hat{N}_1^s)^2\ .
\end{eqnarray}
\end{subequations}

In the early stage, from eqs.(8a)-(8c), the solutions are
\begin{subequations}
\begin{eqnarray}
N^{f-f}&=&f_f^2R_f\theta^3=f_fN\ , \\
N^{f-s}&=&f_ff_sR_f\theta^3=f_sN\ , \\
N^{s-s}&=&rf_s^2R_f\theta^3=rf_s^2f_f^{-1}N\ .
\end{eqnarray}
\end{subequations}These results show that nucleus composition depends only on the flux ratio except slow-slow nuclei, which have much lower density than others, in the early stage, and this is consistent with the results displayed in the inset of Fig. 4(a).

In the late stage, using Eqs.(10) and (12), the solutions are
\begin{subequations}
\begin{eqnarray}
N^{f-f}&\sim&R_f^{-1}R_s^{\frac{2}{3}}f_f^2f_s^{-\frac{2}{3}} \theta^{\frac{1}{3}}=rf_f^2f_s^{-1}N\ , \\
N^{f-s}&\sim&R_s^{-\frac{1}{3}}f_ff_s^{\frac{1}{3}} \theta^{\frac{1}{3}}=f_fN\ , \\
N^{s-s}&\sim&R_s^{-\frac{1}{3}}f_s^{\frac{4}{3}} \theta^{\frac{1}{3}}=f_sN\ .
\end{eqnarray}
\end{subequations}This means that the ratio of the nucleus is $ N^{f-f}:N^{f-s}:N^{s-s}=rf_f^2f_s^{-1}:f_f:f_s $, and therefore $N^{f-f}\ll N^{f-s},N^{s-s}$ when the difference between $ f_{f} $ and $ f_{s} $ is not substantial. This result shows that slow adatoms tend to be used as nuclei more frequently than fast adatoms, and this agrees with the trend shown in Fig. 4(a). The number of s-s nuclei in simulation is not so little as eqs.(14), because it includes nuclei in the early stage like as eqs.(12).

In summary, we have simulated submonolayer epitaxial growth with 2 kinds of adatoms and reached the conclusion that slower adatoms tend to form the nuclei of islands and also determine island density and size distribution. We also conducted a theoretical analysis using the rate equation of the point-island model and confirmed these results. 

 We believe our finding, that the slow adatom determines the nucleation and volume distribution, is universal, and therefore can also be applied to the growth of InAlAs quantum dots.It is worth noting a report that the results for 3D islands show similar properties to those of 2D islands \cite{b13}.

 \end{document}